\documentstyle[aps]{revtex}
\begin{document}
\draft

\title{
        Well-posed forms of the 3+1 conformally-decomposed 
        Einstein equations~\footnote{Work supported by CONICOR, CONICET, SeCyT 
UNC, and NSF Grant No. PHY94-07194.
}
       }

\author{Simonetta Frittelli$^{a}$\thanks{e-mail: 
        simo@orion.physics.duq.edu}
        \and
        Oscar A. Reula$^{b}$\thanks{e-mail: 
        reula@fis.uncor.edu}
}

\address{
        $^a$ Physics Department, Duquesne University, 
                Pittsburgh, PA 15282
                USA\\
        $^b$ FaMAF, Universidad Nacional de C\'{o}rdoba, 
        Ciudad Universitaria, 5000 C\'{o}rdoba, Argentina               
        }

\date{\today}

\maketitle

\begin{abstract}

We show that well-posed, conformally-decomposed formulations of the
3+1 Einstein equations can be obtained by densitizing the lapse and by
combining the constraints with the evolution equations.  We compute
the characteristics structure and 
verify the constraint propagation of these new well-posed
formulations. In these formulations, the trace of the extrinsic
curvature and the determinant of the 3-metric are singled out from the
rest of the dynamical variables, but are evolved as part of the
well-posed evolution system.  The only free functions are the lapse
density and the shift vector.  We find that there is a 3-parameter
freedom in formulating these equations in a well-posed manner, and
that  part of the parameter space found consists of formulations with
causal characteristics, namely, characteristics that lie only within
the lightcone. In particular there is a 1-parameter family of systems
whose characteristics are either 
normal to the slicing or lie  along the lightcone of
the evolving metric.

\end{abstract}
\pacs{PACS}

\section{Introduction} \label{sec:1}


Analytical work in recent years has produced a number of systems
of evolution equations which are equivalent to the Einstein
equations at the constraint manifold, and which have a well posed
initial value
formulation~\cite{bona-masso92,frittelli-reula94,massoletter,york,frittelliletter,helmut96,frittelli-reula99}.

What motivates interest in this type of result is a general
understanding (see for instance~\cite{kreissbook}) that
explicit well-posedness  would be relevant in implementing consistent
and
stable numerical algorithms to integrate blackhole spacetimes.

The well-posed schemes for which a numerical code has been
implemented appear not to exibit significant improvements over
other methods, there being several factors relevant to numerical
implementation which play a significant role.  What is puzzling,
however, is that,  on the other hand, there have been numerical
simulations with apparently better behavior, but which are based on
systems which do not seem to have the well-posed character. One
preponderant feature of these 
numerically more robust schemes is that they are built on a
decomposition of the intrinsic metric into a metric of unit
determinant and the determinant itself, and of the extrinsic
curvature into trace and trace-free part. With slight variations,
this way of evolving the 3+1 Einstein equations has been considered
by~\cite{kurki-laguna-matzner93,shibata}. Quite recently, this form
has been shown to possess striking computational advantages over
the standard form~\cite{BS99}.  We refer to this general scheme as
a conformally-decomposed formulation of the 3+1 Einstein
equations.    

It is difficult to explain the success of these systems as opposed
to the well-posed evolution schemes, or even to the standard (ADM)
evolution schemes.  The relative sizes of the
fields of well-posed systems are roughly the same for different
spectral frequencies in a Fourier representation, which helps
explain the stability of the system via numerical analysis. 
However, in the conformally-decomposed systems this does not happen
in general (for standard norms), as it does not happen for the
standard ADM system, thus making it more difficult to justify their
relative better behavior. 

We can speculate on two features that can possibly bear relevance
to well behaved numerical evolution. One feature is that good
behavior in evolution is related to constraint violations. If the
system preserves more accurately the constraints, then the
evolution remains closer to the constraint submanifold, which
contains the physical solutions. Outside this submanifold the
solutions are unphysical; thus, there is no compelling reason to rule
out fast growths for seemingly tame initial data for unphysical
solutions. Thus, we suggest that controlling the constraint
violations may lead to well-behaved numerical evolution. In this
respect, it has been shown~\cite{lambdavar}, that (at least in the linearized
case) 
there are well-posed modifications of the Einstein equations
outside the constraint submanifold which make that submanifold an
attractor, thus improving the chances
of building
numerical codes with better behaved constraint propagation.
Another possible cause of concern for generating numerical instabilities is the
nature of the boundary conditions which
are ussually imposed. 
The initial-boundary value problem for the Einstein equations has
not been studied for the systems used in numerical simulations and
where instabilities have been found (see nevertheless~\cite{helmutnagy} for a
complete theory of boundary values for Einstein's equations in
conformal  frame variables, and~\cite{stewart98} for a linearized
study), thus the set of boundary conditions for which the
constraint equations are satisfied is in general not known. In
dealing with this problem, establishing well-posedness for the Cauchy
problem is a necessary first
step.

The other feature which could give rise to numerical instabilities is
the relative sizes of the ``longitudinal'' and
the ``radiative'' modes in general relativity. In all non-trivial
asymptotically flat solutions (either vacuum or with matter
satisfying the appropiate energy conditions) the positivity of the
mass implies the existence of longitudinal modes, and there are
many astrophysically relevant cases where there is an approximate
local notion of longitudinal versus transverse modes, and where the
former are several orders of magnitude bigger that the latter. If
they are not properly separated in the numerical algorithms, the
errors caused by finite differencing might be of the order of the
``radiative'' modes, and bad behavior can be expected.  The
separation of the conformal freedom in the more successful codes
can perhaps be thought of as a way of dealing with this issue, or
at least isolating it. 

In this work, we focus on this latter aspect. A technique for taking advantage
of the conformal
factor to  partially decouple the ``longitudinal''  and
``transversal'' modes was used to obtain results on the Newtonian limit
of
general  relativity~\cite{frittelli-reula94}. In that case the
conformal field was fixed via an elliptic equation, decoupling in
this way the more prominent Newtonian potential to first order from
the radiative degrees of freedom. Further studies on this problem
would be critical to obtain realistic simulations of most
astrophysically relevant problems. 

Here we construct 3-parameter families of first-order well-posed systems wich
share some of the properties of the more successful systems, such
as the conformal decomposition of the fundamental fields, in the
hope that their study would help understand what is causing them
to behave better than others. In Section~\ref{sec:2} we apply
techniques similar to those we used
in~\cite{frittelliletter,frittelli-reula99} in order to obtain
versions of the 3+1 equations that are conformally-decomposed but
which are well posed.  Additionally, we calculate the structure of
characteristics and show that for a open region in parameter 
space the resulting equations are metric-causal, namely they have all 
propagation cones inside or coincident with the light cone.
There is even a one parameter subfamily for which propagation is either
along the light cone or normal to the slices.

Furthermore we show that the constraints are propagated by
these well-posed evolution equations.  
As oposed to~\cite{brugman}, where also analytical studies of 
systems with this decomposition have been done,
in this work the trace of the extrisic
curvature and the determinant of the intrinsic metric are
considered dynamical variables and are evolved jointly with the
rest of the system.

\section{System II}\label{sec:2}


The conformally-decomposed system that we 
take as starting point
has appeared in~\cite{BS99}, and is a variation of the system used
by Shibata and Nakamura~\cite{shibata}. It is a system of 15
equations for 15 variables $(\phi, K, \tilde{\gamma}_{ij},
\tilde{A}_{ij}, \tilde{\Gamma}^i)$, and is referred to as System
II in~\cite{BS99}, to distinguish it from the standard 3+1
Einstein equations~\cite{york79}.  These variables are related to
the intrinsic metric $\gamma_{ij}$ and extrinsic curvature
$K_{ij}$ as follows:
\begin{mathletters}
\begin{eqnarray}
        e^{4\phi} &=& \det(\gamma_{ij})^{1/3}\\
        \tilde{\gamma}_{ij} & = & e^{-4\phi}\gamma_{ij}\\
        K  &=& \gamma^{ij}K_{ij}\\
        \tilde{A}_{ij} &=& e^{-4\phi}\left( 
                                K_{ij} 
                  - \frac13 \gamma_{ij}K\right) \\
        \tilde{\Gamma}^i &=& -\tilde{\gamma}^{ij},_j
\end{eqnarray}
\end{mathletters}
where $\tilde{\gamma}^{ij}$ is the inverse of
$\tilde{\gamma}_{ij}$. The Einstein equations in terms of these
variables are equivalent to the following:
\begin{mathletters}\label{bs}
\begin{eqnarray}
  \frac{d}{dt}\phi &=& -\frac16 \alpha K\\
  \frac{d}{dt}\tilde{\gamma}_{ij}& = & -2\alpha\tilde{A}_{ij}\\
  \frac{d}{dt}K                                 &=& 
        -\gamma^{ij}D_iD_j\alpha 
        +\alpha\left(\tilde{A}_{ij}\tilde{A}^{ij}
                     +\frac13 K^2
                \right)
        +\frac12\alpha(\rho+S)\\
  \frac{d}{dt}\tilde{A}_{ij}                    &=& 
        e^{-4\phi}\left( -(D_iD_j\alpha)^{TF}
                        +\alpha(R_{ij}^{TF} - S_{ij}^{TF})
                  \right)
        +\alpha(K\tilde{A}_{ij}-2\tilde{A}_{il}\tilde{A}^l{}_j) \\
  \frac{\partial}{\partial t}\tilde{\Gamma}^i   &=& 
        -2\tilde{A}^{ij}\alpha_j
        +2\alpha\left(\tilde{\Gamma}^i_{jk}\tilde{A}^{kj}
                      -\frac23\tilde{\gamma}^{ij}K,_j
                      -\tilde{\gamma}^{ij}S_j
                      +6\tilde{A}^{ij}\phi,_j
                \right)                         \nonumber\\
                                                &&
        -\frac{\partial}{\partial x^j}
         \left( \beta^l\tilde{\gamma}^{ij},_l
                -2\tilde{\gamma}^{m(j}\beta^{i)},_m
                +\frac23\tilde{\gamma}^{ij}\beta^l,_l
         \right).
\end{eqnarray}
Here $\alpha$ is the lapse function, $\beta^i$ is the shift
vector, and $\tilde{\Gamma}^i_{jk}$ are the connection
coefficients of $\tilde{\gamma}_{ij}$. The superscript $^{TF}$
denotes trace-free part, e.g. $R_{ij}^{TF} = R_{ij} -
\gamma_{ij}R/3$. Indices are raised and lowered with
$\tilde{\gamma}^{ij}$ and its inverse. We use the shorthand
notation
\begin{equation}
\frac{d}{dt} \equiv \frac{\partial}{\partial t} - \pounds_\beta
\end{equation}
where $\pounds_\beta$ is the Lie derivative along $\beta^i$. We
have, as well,
\begin{eqnarray}
R^{TF}_{ij} &=& R_{ij}-\frac13 \gamma^{ij}\gamma^{kl}R_{kl}\\
R_{ij}&=&\tilde{R}_{ij}+R^{\phi}_{ij}\\
R^{\phi}_{ij}&=& -2\tilde{D}_i\tilde{D}_j\phi 
                 -2\tilde{\gamma}_{ij}
                   \tilde{\gamma}^{kl}\tilde{D}_k\tilde{D}_l\phi
                 +4 \tilde{D}_i\phi\tilde{D}_j\phi
                 -4\tilde{\gamma}_{ij}\tilde{\gamma}^{kl}
                        \tilde{D}_k\phi\tilde{D}_l\phi\\
\tilde{R}_{ij} &=& 
        -\frac12\tilde{\gamma}^{kl}\tilde{\gamma}_{ij,kl}
        +\tilde{\gamma}_{k(i}\tilde{\Gamma}^k{},_{j)}
                                                \nonumber\\
                                                &&
        +\tilde{\Gamma}^k\tilde{\Gamma}_{(ij)k}
        +\tilde{\gamma}^{lm}\left(
        2\tilde{\Gamma}^k{}_{l(i}\tilde{\Gamma}_{j)km}
        +\tilde{\Gamma}^k{}_{il}\tilde{\Gamma}_{kmj}
                                      \right)           \\
\tilde{\Gamma}^k{}_{ij} &=&
        \frac12\tilde{\gamma}^{kl}\left(
        \tilde{\gamma}_{il,j}
        +\tilde{\gamma}_{jl,i}
        -\tilde{\gamma}_{ij,l}   \right).
\end{eqnarray}
\end{mathletters}
This system is first-order in time and second-order in space, thus
it is of second order overall. We show how System II can be
handled in order to be turned into a well-posed form.  Firstly we
reduce the system to a straightforward first order form, and
subsequently we densitize the lapse and combine the constraints
into the evolution equations.

\subsection{System II reduced to first-order form}

We define a set of 12 additional variables 
\begin{mathletters}\label{new}
\begin{equation} 
        V_{ijk} 
   \equiv 
        \tilde{\gamma}_{ij,k}
        -\frac35 \tilde{\gamma}_{k(i}\tilde{\gamma}_{j)n,s}
                 \tilde{\gamma}^{ns}
        +\frac15 \tilde{\gamma}_{ij}\tilde{\gamma}_{kn,s}
                 \tilde{\gamma}^{ns}, 
\end{equation}
which is trace free in all its indices, namely:
$V_{ijk}\tilde{\gamma}^{ij}=0$ and $V_{ijk}\tilde{\gamma}^{jk}=
V_{ijk}\tilde{\gamma}^{ik}=0$,  and another set of 3 additional
variables 
\begin{equation} Q_i \equiv \phi,_i.
\end{equation} 
\end{mathletters}

Evolution equations for these new variables are obtained by
taking a time derivative of (\ref{new}) and commuting time and
spatial derivatives in the resulting right-hand sides.  The
complete system of equations is now

\begin{mathletters}\label{firstbs}
\begin{eqnarray}
  \frac{d}{dt}\phi &=& -\frac16 \alpha K\\
  \frac{d}{dt}\tilde{\gamma}_{ij}& = & -2\alpha\tilde{A}_{ij}\\
  \frac{d}{dt}K                                 &=& 
        -\gamma^{ij}D_iD_j\alpha 
        +\alpha\left(\tilde{A}_{ij}\tilde{A}^{ij}
                     +\frac13 K^2
                \right)
        +\frac12\alpha(\rho+S)\\
  \frac{d}{dt}\tilde{A}_{ij}                    &=& 
        e^{-4\phi}\left( -(D_iD_j\alpha)^{TF}
                        +\alpha(R_{ij}^{TF} - S_{ij}^{TF})
                  \right)
        +\alpha(K\tilde{A}_{ij}
        -2\tilde{A}_{il}\tilde{A}^l{}_j) \\
  \dot{\tilde{\Gamma}^i} -\beta^l\tilde{\Gamma}^i,_l    &=&     
        -2\tilde{A}^{ij}\alpha,_j
        +2\alpha\left(\tilde{\Gamma}^i{}_{jk}\tilde{A}^{kj}
                      -\frac23\tilde{\gamma}^{ij}K,_j
                      -\tilde{\gamma}^{ij}S_j
                      +6\tilde{A}^{ij}Q_j
                \right)
        -\beta^l,_j \tilde{\gamma}^{ij},_l      \nonumber\\
                                                        &&
        +\tilde{\Gamma}^m\beta^i,_m
        +\tilde{\gamma}^{mi},_j\beta^j,_m
        +2\tilde{\gamma}^{m(i}\beta^{j)},_{mj}
        +\frac23\tilde{\Gamma}^i\beta^l,_l
        +\frac23\tilde{\gamma}^{ij}\beta^l,_{lj}                                
                                                                       \\
   \dot{V}_{ijk}-\beta^l V_{ijk,l}      &=&     
        -2\alpha \tilde{A}_{ij,k}
        +\frac65\alpha\tilde{\gamma}_{k(i}
         \tilde{A}_{j)m,n}\tilde{\gamma}^{mn}
        -\frac25\alpha\tilde{\gamma}_{ij}
         \tilde{A}_{km,n}\tilde{\gamma}^{mn}
                                                \nonumber\\
                                                        &&
        +\beta^l,_kV_{ijl}
        +\beta^l,_i V_{ljk}
        +\beta^l,_j V_{ilk}
        -2\alpha,_k\tilde{A}_{ij}
                                                \\
                                                &&
        +\frac65\tilde{\gamma}_{k(i}\tilde{A}_{j)}^n
         \alpha,_n
        +\frac65\alpha\tilde{A}_{k(i}\tilde{\Gamma}_{j)}
        -\frac25\alpha\tilde{A}_{ij}\tilde{\Gamma}_k
                                                \\
                                                &&
        +\tilde{\gamma}_{l(i}\beta^l,_{j)k}
        -\frac35\beta^l,_{l(j}\tilde{\gamma}_{i)k}
        +\frac15\tilde{\gamma}_{ij}\beta^l,_{lk}
                                                \\
                                                &&
        -\beta^l,_{ns}\tilde{\gamma}^{ns}
        \left(
        \frac35\tilde{\gamma}_{k(i}\tilde{\gamma}_{j)l}
        -\frac15\tilde{\gamma}_{ij}\tilde{\gamma}_{kl}
        \right)
                                                \\
                                                &&
        +\left(2\alpha\tilde{A}^{ns}
              +\tilde{\gamma}^{ns}\beta^s,_m
        \right)
        \left(\frac35\tilde{\gamma}_{k(i}
                     \tilde{\gamma}_{j)n,s}
             -\frac15\tilde{\gamma}_{ij}
                     \tilde{\gamma}_{ks,n}
        \right)
                                                \\
   \dot{Q}_i - \beta^l Q_{i,l} &=& 
        -\frac16 \alpha K,_i
        +\beta^l,_i Q_l
        -\frac16 \alpha,_i K
\end{eqnarray}
\end{mathletters}
where, as before, $R^{TF}_{ij} = R_{ij}-\frac13
\gamma^{ij} \gamma^{kl} R_{kl}$, with
$R_{ij} = \tilde{R}_{ij} + R^{\phi}_{ij}$, and 
\begin{mathletters}
\begin{eqnarray}
      R^{\phi}_{ij} 
  &=& 
        -2\tilde{D}_iQ_j 
        -2\tilde{\gamma}_{ij}
          \tilde{\gamma}^{kl}\tilde{D}_kQ_l
        +4 Q_iQ_j
        -4\tilde{\gamma}_{ij}\tilde{\gamma}^{kl}Q_kQ_l,\\
      \tilde{R}_{ij} 
   &=& 
        -\frac12 V_{ijk},^k
        +\frac{7}{10}\tilde{\gamma}_{k(i}
                     \tilde{\Gamma}^k{},_{j)}
        +\frac{1}{10}\tilde{\gamma}_{ij}
                     \tilde{\Gamma}^k,_k
                                                \nonumber\\
                                                &&
        -\frac{3}{10}\left(
                 \tilde{\Gamma}^kV_{k(ij)}
                +\tilde{\Gamma}_{(i}V_{j)k}{}^k
                +\frac{9}{10}
                 \tilde{\Gamma}_i\tilde{\Gamma}_j
                     \right)
                                                \nonumber\\
                                                &&
        -\frac15\left(
                 V_{ijk}\tilde{\Gamma}^k
                +\tilde{\gamma}_{ij}
                    \tilde{\Gamma}^kV_{km}{}^m
                -\frac{1}{10}
                    \tilde{\gamma}_{ij}
                    \tilde{\Gamma}^k\tilde{\Gamma}_k
                \right)
                                                \nonumber\\
                                                &&
                  +\tilde{\Gamma}^k\tilde{\Gamma}_{(ij)k}
                  +2\tilde{\Gamma}^{kl}{}_{(i}
                    \tilde{\Gamma}_{j)kl}
                  +\tilde{\Gamma}^k{}_{il}
                   \tilde{\Gamma}_{kj}{}^l,
                                                        \\
\tilde{\Gamma}^k{}_{ij} &=&
                 V^k{}_{(ij)}
        -\frac12 V_{ij}{}^k
        -\frac15\delta^k_{(i}\tilde{\Gamma}_{j)}
        +\frac25\tilde{\gamma}_{ij}\tilde{\Gamma}^k,
\end{eqnarray}
\end{mathletters}
and indices are raised and lowered with $\tilde{\gamma}^{ij}$ and
$\tilde{\gamma}_{ij}$ respectively. The derivatives of the form 
$\tilde{\gamma}_{ij,k}$ that appear in the right-hand sides of 
(\ref{firstbs}) must be interpreted simply as shorthands for
combinations of the fields $\tilde{\Gamma}^i$ and $V_{ijk}$, via
\begin{equation}
        \tilde{\gamma}_{ij,k}
   =
        V_{ijk}
        +\frac35\tilde{\gamma}_{k(i}\tilde{\Gamma}_{j)}
        -\frac15\tilde{\gamma}_{ij}\tilde{\Gamma}_k
\end{equation}
For this first-order system to be equivalent to the Einstein
equations (in the sense that its set of solutions is the same as
that of the Einstein equations), the following set of constraints
must be imposed on the initial data (and are subsequently
preserved by the evolution, as will be shown in the next section):
\begin{mathletters}\label{const}
\begin{eqnarray}
   {\cal H} &=& 
        \gamma^{ij}R_{ij} - 
        \tilde{A}_{ij} \tilde{A}^{ij}
        +\frac23 K^2
        -2\rho                  \label{hamilt}  \\
   {\cal P}_i &=&
        \tilde{\gamma}^{jl}D_l\tilde{A}_{ij}
        -\frac23 D_i K
        +4Q_l\tilde{A}^l{}_i
        +\frac43 KQ_i
        -S_i                    \label{moment}\\
   {\cal G}^i &=&
        \tilde{\Gamma}^i 
        +\tilde{\gamma}^{ij},_j  \label{gamma}  \\
   {\cal Q}_i &=&
        Q_i - \phi,_i           \label{q}\\
   {\cal V}_{ijk} &=&
          V_{ijk} 
        - \tilde{\gamma}_{ij,k} 
        +\frac35 \tilde{\gamma}_{k(i}
                 \tilde{\gamma}_{j)n,s}
                 \tilde{\gamma}^{ns}
        -\frac15 \tilde{\gamma}_{ij}
                 \tilde{\gamma}_{kn,s}
                 \tilde{\gamma}^{ns},\label{v}
\end{eqnarray}
where
\begin{eqnarray}
     \gamma^{ij}R_{ij}
   &=&
      e^{-4\phi}\left(
          \tilde{\Gamma}^l,_l
        -8\tilde{D}^lQ_l
        -Q^lQ_l 
        -\frac12 V_{ijl}V^{ijl}
        -\frac{15}{10}\tilde{\Gamma}^kV_{km}{}^m
        -\frac15 V^m{}_mk\tilde{\Gamma}^k
                \right.                         \\
                                                &&
                \left.
        -\frac{21}{100}\tilde{\Gamma}^k\tilde{\Gamma}_k
        +\tilde{\Gamma}^k\tilde{\Gamma}^m{}_{mk}
        +2\tilde{\Gamma}^{klm}\tilde{\Gamma}_{mkl}
        +\tilde{\Gamma}^{mkl}\tilde{\Gamma}_{mkl}
                \right).
\end{eqnarray}  
\end{mathletters}
Constraints (\ref{hamilt}) and (\ref{moment}) are the hamiltonian
and momentum constraints of the 3+1 decomposition of the Einstein
equations, written in our choice of variables.  Constraints
({\ref{gamma}), (\ref{q}) and (\ref{v}) arise in turning the
original second-order system into first order. 

In (\ref{firstbs}) and (\ref{const}), the derivative $D_l$ is the
covariant derivative with respect to $\gamma_{ij}$, and is related
to $\tilde D_l$ by undifferentiated terms:
\begin{equation}
        \Gamma^k{}_{ij}
   =
        \tilde{\Gamma}^k{}_{ij} 
        +2\left(
                 Q_i\delta^k_j
                +Q_j\delta^k_i
                -Q_l\tilde{\gamma}^{kl}\tilde{\gamma}_{ij}
          \right)
\end{equation}

\subsection{Taking advantage of the available freedom}

In this section we take advantage of two facts that have been used
successfully in similar
problems~\cite{choquet-ruggeri83,frittelliletter,frittelli-reula99}. 
Firstly, we densitize the lapse $\alpha$ (and in doing so we
introduce a free function, referred as ``slicing density''
in~\cite{york98}):
\begin{equation}
        \alpha=e^{4a\phi}\sigma
\end{equation}
Like the shift vector $\beta^i$, the lapse density $\sigma$ will
be considered arbitrary but fixed, a source function independent
of the dynamical fields. 

Secondly, the evolution equations can be combined with the
constraints without altering the set of solutions.  We add the
scalar constraint with a factor $b$ to the evolution equation for
$K$ and we add the vector constraint to the evolution equations
for $\tilde{\Gamma}^i$ and $Q_i$, with factors of $c$ and $d$
respectively. In this manner we obtain a system of the form
\begin{equation}
       \dot{u} = \mbox{\boldmath $A$}^i(u) \nabla_i u + B(u).
\end{equation}
A system of this form is known to be well posed if the
matrix-valued vector $\mbox{\boldmath $A$}^i(u)$ admits a
symmetrizer, namely, a positive definite, symmetric, bi-linear
form \mbox{\boldmath $H$}, in the space of the fields $u$, whose
product with $\mbox{\boldmath $A$}^i(u)$ yields a
symmetric-bilinear-form-valued vector.  Thus, in order to
determine well-posedness, it suffices to consider the principal
part of the system. 

In this case, the principal terms are
\begin{mathletters}\label{combs}
\begin{eqnarray}
  \dot{\phi} 
     &=& 
        \beta^l\phi,_l \\
  \dot{\tilde{\gamma}}_{ij} 
     &=& 
        \beta^l\tilde{\gamma}_{ij,l}                    \\
  \dot{K}                               
     &=& 
        \beta^l K,_l
        -\alpha(4a+8b)e^{-4\phi}\tilde{\gamma}^{kl}Q_{k,l} 
        +\alpha b e^{-4\phi}\tilde{\Gamma}^l{},_l                       \\
  \dot{\tilde{A}}_{ij}                  
     &=& 
        \beta^l\tilde{A}_{ij,l}
        +e^{-4\phi}\alpha\left(
        -\frac12 \tilde{\gamma}^{kl}V_{ijk,l}
        +\frac{7}{10}\left(
                 \tilde{\gamma}_{k(i}\tilde{\Gamma}^k,_{j)}
         -\frac13\tilde{\gamma}_{ij}\tilde{\Gamma}^k,_k
                     \right)
                         \right)                \nonumber\\
                                                        &&
        -2(2a+1)e^{-4\phi}\alpha\left(
                Q_{(i,j)}
                -\frac13\tilde{\gamma}_{ij}
                        \tilde{\gamma}^{kl}Q_{k,l}
                \right) 
                                                \\
  \dot{\tilde{\Gamma}}^i        
     &=& 
        \beta^l\tilde{\Gamma}^i,_l
        +\alpha c\tilde{A}^{il},_l
        -\frac23(c+2)\alpha \tilde{\gamma}^{il} K,_l    \\
   \dot{V}_{ijk}        
     &=&
        \beta^l V_{ijk,l}
        -2\alpha \tilde{A}_{ij,k}
        +\frac65\alpha\tilde{\gamma}_{k(i}
                      \tilde{A}_{j)m,n}
                      \tilde{\gamma}^{mn}
        -\frac25\alpha\tilde{\gamma}_{ij}
                      \tilde{A}_{km,n}
                      \tilde{\gamma}^{mn}                               \\
   \dot{Q}_i 
      &=& 
        \beta^l Q_{i,l}
        -\frac{\alpha}{6}(1+4d)K,_i
        +d \alpha\tilde{\gamma}^{jl}\tilde{A}_{ij,l}.
\end{eqnarray}
\end{mathletters}
Our aim is to show that there exist choices of the numerical
factors $a,b,c,d$ such that the system (\ref{combs}) is
symmetrizable, therefore, well posed. We establish this result by
defining a candidate symmetrizer \mbox{\boldmath$H$} given as
\begin{eqnarray}\label{energy}
  \bar u \mbox{\boldmath$H$} u 
    &=&
        \phi^2 + 
        \delta^{ik}\delta^{jl}\tilde{\gamma}_{ij}
                              \tilde{\gamma}_{kl}
        +n_1^2 e^{-4\phi} K^2
        +\tilde{A}^{ij}\tilde{A}_{ij}
        +n_2^2 e^{-4\phi}\tilde{\Gamma}^i\tilde{\Gamma}_i
        +\frac{e^{-4\phi}}{4}V_{ijk}V^{ijk}             
                                                \nonumber\\
                                                        &&
        +n_3^2 e^{-4\phi} Q^iQ_i
\end{eqnarray}
where $n_1, n_2, n_3$ are any fixed real numbers different from
zero and bounded, so that  $C^{-1} \mbox{\boldmath$I$}\le
\mbox{\boldmath$H$}  \le C\mbox{\boldmath$I$}$  where $C$ is a
positive constant and \mbox{\boldmath$I$} is the identity operator
on the space of $u$ . 

We can easily arrange the values of $a,b,c,d$ so that
symmetry of $\mbox{\boldmath$HA$}^i(u)$ is attained.  To this
effect, they must satisfy
\begin{mathletters}\label{ens}
\begin{eqnarray}
        \frac{7}{10}   & = & n_2^2 c\\ 
        -2(2a+1)       & = & n_3^2 d\\
        -(4a+8b) n_1^2 & = & -\frac16(1+4d)n_3^2\\
        n_1^2 b        & = & -\frac23 n_2^2 (c+2)
\end{eqnarray}
\end{mathletters}
There is clearly plenty of freedom in the choice of $a,b,c,d$,
since any choice that results in non-vanishing $n_1,n_2,n_3$ is
allowed.  The freedom is thus parametrized by the values of
$n_1^2, n_2^2, n_3^2$, since these can take independent positive
values. Thus our four parameters $a,b,c,d$ are not all
independent, but there is a relationship between them that reduces
the freedom to 3 independent parameters. We can solve (\ref{ens})
for $a,b,c,d$ in terms of $n_1, n_2, n_3$, which yields:
\begin{mathletters}\label{parametric}
\begin{eqnarray}
        a&=& \frac{9/5+8n_2^2+n_3^2/8}{(3n_1^2+2)}\\
        b&=& -\frac{2}{3n_1^2}\left(\frac{7}{10}+2n_2^2\right)\\
        c&=&\frac{7}{10n_2^2}   \\
        d&=&-\frac{2}{n_3^2}
             \left(2\frac{9/5+8n_2^2+n_3^2/8}{(3n_1^2+2)}
                   +1\right)
\end{eqnarray}
\end{mathletters}
It is clear from (\ref{parametric}) that $a$ and $c$ will take
only strictly positive values, and $b$ and $d$ will take only
strictly negative values, for all real values of $n_1, n_2, n_3$
different from zero.

\subsection{Structure of characteristics}

The system (\ref{combs}) is of the form
\begin{equation}
 \mbox{\boldmath$A^a$}\frac{\partial u}{\partial x^a} = 0.
\end{equation}
The characteristic covectors are covectors $\xi_a=(\xi_i,-v)$ such
that $\xi_i\xi_j\gamma^{ij}=1$ and such that
\begin{equation}\label{det}
      \det(\mbox{\boldmath$A^a$}\xi_a)=0
\end{equation}
The values of $v$ that satisfy (\ref{det}) for every direction
$\xi_i$ are the characteristic speeds in that direction.  In order
to find these values we set up an eigenvalue problem for the
principal symbol $\mbox{\boldmath$A^a$}\xi_a$ and find the null
eigenvectors. The eigenvalue problem is
\begin{mathletters}\label{eigencombs}
\begin{eqnarray}
  n^a\xi_a\phi &=& 0                    \label{eigencombs1}\\
  n^a\xi_a\tilde{\gamma}_{ij}& = & 0    \label{eigencombs2}\\
  n^a\xi_a K                            
     &=& 
        -(4a+8b)e^{-4\phi}\xi^kQ_{k} 
        +  b e^{-4\phi}\xi_l\tilde{\Gamma}^l    
                                        \label{eigencombs3}\\
  n^a\xi_a\tilde{A}_{ij}                        
     &=& 
        e^{-4\phi}\left(
        -\frac12 \xi^kV_{ijk}
        +\frac{7}{10}\left(
        \tilde{\gamma}_{k(i}\xi_{j)}\tilde{\Gamma}^k
        -\frac13\tilde{\gamma}_{ij}\xi_k\tilde{\Gamma}^k
                     \right)
                \right)
                                \nonumber\\
                                                &&
                -2(2a+1)e^{-4\phi}\left(
                \xi_{(j}Q_{i)}
                -\frac13\tilde{\gamma}_{ij}\xi^kQ_k
                        \right) 
                                \label{eigencombs4}\\
  n^a\xi_a\tilde{\Gamma}^i      
     &=& 
         c\xi_l\tilde{A}^{il}
        -\frac23(c +2) \xi^i K          \label{eigencombs5}\\
   n^a\xi_a V_{ijk}     
     &=&
        -2\xi_k\tilde{A}_{ij}
        +\frac65\tilde{\gamma}_{k(i}\tilde{A}_{j)m}\xi^m
        -\frac25\tilde{\gamma}_{ij}\xi^m\tilde{A}_{km}          
                                        \label{eigencombs6}\\
   n^a\xi_a Q_i 
     &=& 
        -\frac16(1+4d)\xi_iK
        +d \xi^j\tilde{A}_{ij}    \label{eigencombs7}
\end{eqnarray}
\end{mathletters}
Clearly, $n^a\xi_a=0$ allows for 18 eigenvectors.  This is because
(\ref{eigencombs5}), (\ref{eigencombs6}) and (\ref{eigencombs7})
in this case constitute an overdetermined system of 18 homogeneous
equations for 6 unknowns $(\tilde{A}_{ij}, K)$, with zero as the
only solution, whereas (\ref{eigencombs3}) and (\ref{eigencombs4})
constitute a system of 6 equations for 18 variables, which leaves
out 12 of the 18 fields $(V_{ijk}, \tilde{\Gamma}^i, Q)$ free. 
Lastly, (\ref{eigencombs1}) and (\ref{eigencombs2}) leave the 6
variables $(\tilde{\gamma}_{ij}, \phi)$ free.  If we represent
the eigenvectors in the form
\begin{equation}
 (\tilde{\gamma}_{ij}, \phi, Q_i, \tilde{\Gamma}^{(L)},
\tilde{\Gamma}^{(T)}_i, V^{(L)}_{ij}, 
V^{(T)}_{ijk}, K, \tilde{A}^{(LL)}, \tilde{A}^{(LT)}_i, \tilde{A}^{(TT)}_{ij})
\end{equation}
where
$
\tilde{\Gamma}^{(L)}\!:= \tilde{\Gamma}^i\xi_i,\;
\tilde{\Gamma}^{(T)}_i\!:=\tilde{\Gamma}_i-
                        e^{-4\phi} \xi_i\tilde{\Gamma}^k\xi_k,\;
V^{(L)}_{ij}\!:= V_{ijk}\xi^k,
V^{(T)}_{ijk}\! := V_{ijk}- e^{-4\phi}\xi_k V_{ijl}\xi^l,\;
\tilde{A}^{(LL)}\! := \tilde{A}^{ij}\xi_i\xi_j,\;
\tilde{A}^{(LT)}_i\!:= \tilde{A}_{ij}\xi^j 
                        -e^{-4\phi}\xi_i \tilde{A}^{kl}\xi_k\xi_l       ,\;
\tilde{A}^{(TT)}_{ij}\! := \tilde{A}_{ij}
                        -2e^{-4\phi}\xi_{(i}\tilde{A}_{j)l}\xi^l
                        +e^{-8\phi}\xi_i\xi_j\tilde{A}^{kl}\xi_k\xi_l,\;
$
then we have 5 eigenvectors corresponding to the five components of the
conformal metric
\begin{mathletters}
\begin{equation} 
        (\tilde{\gamma}_{ij}, 0,0,0,0,0,0,0,0,0);
\end{equation}
we have the determinant as an eigenfield
\begin{equation}
        (0, \phi,0,0,0,0,0,0,0,0);
\end{equation}
we have 7 eigenvectors corresponding to the seven transverse components of
$V_{ijk}$
\begin{equation}
        (0,0,0,0,0,0,V^{(T)}_{ijk},0,0,0);
\end{equation}
we have 3 eigenvectors corresponding essentially to the three components of
$Q_i$
\begin{equation}
        (0,0,Q_i,\frac{4a+8b}{b}\xi^iQ_i,0,
         -2(2a+1)\xi_{(i}Q_{j)} -\frac13\tilde{\gamma}_{ij}
          \left(\frac{7(4a+8b)}{10b}-2(2a+1)\right)Q_l\xi^l,0,0,0,0);
\end{equation} 
and we have 2 eigenvectors corresponding essentially to the two components of
the transverse part of $\tilde{\Gamma}_i$
\begin{equation}
        (0,0,0,0,\tilde{\Gamma}^{(T)}_i,
         \frac{7}{10}\xi_{(i}\tilde{\Gamma}^{(T)}_{j)},0,0,0,0);
\end{equation} 
\end{mathletters}                       
If $n^a\xi_a\neq 0$, then $\tilde{\gamma}_{ij}=\phi=0$, and we can
solve (\ref{eigencombs5}), (\ref{eigencombs6}) and
(\ref{eigencombs7}) for $(V_{ijk}, \tilde{\Gamma}^i, Q)$ in terms
of $\xi_a$ and $(\tilde{A}_{ij}, K)$.  We can substitute
$(V_{ijk}, \tilde{\Gamma}^i, Q)$ into (\ref{eigencombs3}) and
(\ref{eigencombs4}), obtaining thus a system of 6 equations for
the 6 variables $(\tilde{A}_{ij}, K)$ as follows
\begin{mathletters}\label{red}
\begin{eqnarray}
    0
     &=&
        K e^{4\phi}\left(
                 (n^a\xi_a)^2 
                -\frac16(4a+8b)(1+4d)
                +\frac23 b(c+2)
                   \right)
        +\xi\!\cdot\!\tilde{A}\!\cdot\!\xi
        \left( (4a+8b)d-bc \right)              \label{red1}\\
    0
     &=&
        \tilde{A}_{ij} e^{4\phi}\left(1-(n^a\xi_a)^2\right)
        -\frac{K}{3}\left( \frac{7}{5}(c+2)
                           -(2a+1)(1+4d)
                    \right)
        \left(  \xi_i\xi_j
                -\frac13e^{4\phi}\tilde{\gamma}_{ij}
        \right)                                 \nonumber\\
                                                        &&       
        +\left(  \frac{7c}{10}
                -\frac35
                -2d(2a+1)
        \right)
        \left(
                \xi^l\tilde{A}_{l(i}\xi_{j)}                    
                -\frac13\tilde{\gamma}_{ij}
                \xi\!\cdot\!\tilde{A}\!\cdot\!\xi
        \right)                                 \label{red2}
\end{eqnarray}
\end{mathletters}
where we have used the notation
\begin{equation}
   \xi\!\cdot\!\tilde{A}\!\cdot\!\xi
   :=
        \xi_i\tilde{A}^{ij}\xi_j.
\end{equation}
If $1-(n^a\xi_a)^2=0$, then
$K=\xi\!\cdot\!\tilde{A}\!\cdot\!\xi=0$ by (\ref{red1}), which
implies $\xi^l\tilde{A}_{li}=0$ by (\ref{red2}).  However, two of
the five components of $\tilde{A}_{ij}$ are thus free, which means
that there are 4 eigenvectors, essentially labeled by the transverse
components of $\tilde{A}_{ij}$.  We have 2 eigenvectors for $n^a\xi_a=1$
\begin{mathletters}
\begin{equation}
        (0,0,0,0,0,
         -2\tilde{A}^{(TT)}_{ij},0,0,0,0,\tilde{A}^{(TT)}_{ij});
\end{equation} 
and 2 eigenvectors for $n^a\xi_a=-1$
\begin{equation}
        (0,0,0,0,0,
         2\tilde{A}^{(TT)}_{ij},0,0,0,0,\tilde{A}^{(TT)}_{ij}).
\end{equation} 
\end{mathletters}
If $1-(n^a\xi_a)^2\neq 0$, then contracting (\ref{red2}) with
$\xi^i$ yields
\begin{eqnarray}        \label{rered}
0&=&e^{4\phi}\xi^l\tilde{A}_{lj}
   \left(1
         +\frac12\left(\frac{7c}{10}
                      -\frac35
                      -2d(2a+1)
                 \right)
         -(n^a\xi_a)^2
   \right)                              \nonumber\\
                                        &&
 -\frac29e^{4\phi} K\xi_j
        \left(\frac{7}{5}(c+2)-
              (2a+1)(1+4d)
        \right)                         \nonumber\\
                                        &&
 +\frac16\xi\!\cdot\!\tilde{A}\!\cdot\!\xi \;\xi_j
   \left(\frac{7}{10}c-\frac35
        -2d(2a+1)\right)
\end{eqnarray}
Thus, if $1+\frac12(7c/10-3/5-2d(2a+1))-(n^a\xi_a)^2=0$, then
$K=\xi\!\cdot\!\tilde{A}\!\cdot\!\xi=0$ by (\ref{red1}), which
implies that (\ref{rered}) is identically satisfied, thus three
out of the five equations (\ref{red2}) are identities, the
remaining two determining two components of $\tilde{A}_{ij}$. 
Thus two of the five components of $\tilde{A}_{ij}$ are free,
which means that there are 4 eigenvectors, essentially labeled by
the two longitudinal-transverse components of $\tilde{A}_{ij}$.
We have 2 eigenvectors for $n^a\xi_a = \sqrt{(3/5-7c/10-2d(2a+1))/2}$, namely
\begin{mathletters}
\begin{equation}
        (0,0,\frac{d}{C_1}\tilde{A}^{(LT)}_i,0,
        \frac{c}{C_1}\tilde{A}^{(LT)}_i,
         -\frac{4}{5 C_1}\xi_{(i}\tilde{A}^{(LT)}_{j)},
        \frac{6}{5 C_1}
        \left(\tilde{\gamma}_{k(i}\tilde{A}^{(LT)}_{j)}
              -\xi_k\xi_{(i}\tilde{A}^{(LT)}_{j)}
              -\frac13\tilde{\gamma}_{ij}\tilde{A}^{(LT)}_k
        \right),
        0,0,\tilde{A}^{(LT)}_i,
        0),
\end{equation} 
and 2 eigenvectors for $n^a\xi_a = -\sqrt{(3/5-7c/10-2d(2a+1))/2}$, namely
\begin{equation}
        (0,0,-\frac{d}{C_1}\tilde{A}^{(LT)}_i,0,
        -\frac{c}{C_1}\tilde{A}^{(LT)}_i,
         \frac{4}{5 C_1}\xi_{(i}\tilde{A}^{(LT)}_{j)},
        -\frac{6}{5 C_1}
        \left(\tilde{\gamma}_{k(i}\tilde{A}^{(LT)}_{j)}
              -\xi_k\xi_{(i}\tilde{A}^{(LT)}_{j)}
              -\frac13\tilde{\gamma}_{ij}\tilde{A}^{(LT)}_k
        \right),
        0,0,\tilde{A}^{(LT)}_i,
        0),
\end{equation} 
\end{mathletters}
where we have used the shorthand notation
\begin{equation}
   C_1 :=\sqrt{(3/5-7c/10-2d(2a+1))/2}
\end{equation}

But if $1+\frac12(7c/10-3/5-2d(2a+1))-(n^a\xi_a)^2\neq 0$, then
$\xi^l\tilde{A}_{lj}$ is determined by the values of $K$ and
$\xi\!\cdot\!\tilde{A}\!\cdot\!\xi$ by (\ref{rered}), and if
plugged back into (\ref{red2}) it follows that all the components
of $\tilde{A}_{ij}$ are determined by $K$ and
$\xi\!\cdot\!\tilde{A}\!\cdot\!\xi$.  Therefore it is necessary
that $K$ and $\xi\!\cdot\!\tilde{A}\!\cdot\!\xi$ be nonvanishing. 
Contracting (\ref{rered}) with $\xi^j$ we obtain
\begin{eqnarray}\label{rerered}
   0
    &=&
        -\frac29 e^{4\phi}K\left(\frac{7}{5}(c+2)
                        -(2a+1)(1+4d)
                  \right)                       \nonumber\\
                                                        &&
        +\xi\!\cdot\!\tilde{A}\!\cdot\!\xi
   \left(1
         +\frac23\left(\frac{7}{10}c-\frac35-2(2a+1)d
                 \right)
         -(n^a\xi_a)^2
   \right)
\end{eqnarray}
Equations (\ref{red1}) and (\ref{rerered}) form a system of two
homogeneous equations for $K$ and
$\xi\!\cdot\!\tilde{A}\!\cdot\!\xi$.  Thus, for $K$ and
$\xi\!\cdot\!\tilde{A}\!\cdot\!\xi$ to be nonvanishing, it is
necessary that the determinant of the system be zero. The
determinant is 
\begin{equation}\label{chardet}
        \frac{1}{45}
        (-3(n^a\xi_a)^2+2a)(15(n^a\xi_a)^2-9+10bc-80bd+20d-7c)
\end{equation}
It can be seen that, because $a$ and $c$ are strictly positive and
$b$ and $d$ are strictly negative, the four roots of the
determinant are real. For the roots $n^a\xi_a$ of the determinant, we
have
\begin{mathletters}
\begin{eqnarray}
   K
    &=&  \frac92
        \frac{1 -(n^a\xi_a)^2 
                       +\frac23
                      \left(\frac{7c}{10}-\frac35-2d(2a+1)
                      \right)
                }
             {\frac75(c+2)-(2a+1)(1+4d))}
        e^{-4\phi}\xi\!\cdot\!\tilde{A}\!\cdot\!\xi     \\
   Q_i
    &=&
        \left(   d
                -\frac{3(1+4d)}{4}
                \frac{1 -(n^a\xi_a)^2 
                       +\frac23
                      \left(\frac{7c}{10}-\frac35-2d(2a+1)
                      \right)
                        }
                        {\frac75(c+2)-(2a+1)(1+4d))}
        \right)
        \frac{e^{-4\phi}}{n^a\xi_a}
        \xi_i\xi\!\cdot\!\tilde{A}\!\cdot\!\xi\\
   \tilde{\Gamma}^{(L)}
    &=&
        \left(   c
                -3(c+2)\frac{1 -(n^a\xi_a)^2 
                       +\frac23
                      \left(\frac{7c}{10}-\frac35-2d(2a+1)
                      \right)
                }
             {\frac75(c+2)-(2a+1)(1+4d))}
        \right)\frac{\xi\!\cdot\!\tilde{A}\!\cdot\!\xi}
                    {n^a\xi_a}                          \\
   \tilde{\Gamma}^{(T)}_i 
    &=& 0                                               \\
   V^{(L)}_{ij}
    &=&
        -\frac{9}{5n^a\xi_a}
        \left( e^{-4\phi}\xi_i\xi_j
              -\frac13\tilde{\gamma}_{ij}
        \right)\xi\!\cdot\!\tilde{A}\!\cdot\!\xi        \\
   V^{(T)}_{ijk}
   &=&
        -\frac{6}{5n^a\xi_a}
        \left( \xi_i\xi_j
              -\tilde{\gamma}_{k(i}\xi_{j)}
        \right)e^{-4\phi}\xi\!\cdot\!\tilde{A}\!\cdot\!\xi\\
   \tilde{A}^{(LT)}_i
    &=&
        0                                               \\
   \tilde{A}^{(TT)}_{ij}
    &=&
        \frac{e^{-4\phi}}{2}
        (e^{-4\phi}\xi_i\xi_j-\tilde{\gamma}_{ij})
        \xi\!\cdot\!\tilde{A}\!\cdot\!\xi
\end{eqnarray}
\end{mathletters}
These are clearly four distinct eigenvectors, since there are four
distinct values of $n^a\xi_a$ given by the roots of the determinant
(\ref{chardet}). These can be thought as being labeled, essentially, by $K$
or $\xi\!\cdot\!\tilde{A}\!\cdot\!\xi$ indistinctly, or we can associate
one characteristic speed to $K$ and the other one to
$\xi\!\cdot\!\tilde{A}\!\cdot\!\xi$, as we prefer to do below. 

Summarizing, we have characteristic speeds
obtained from the following distinct values of $n^a\xi_a$:
\begin{description}
\item[a)] $n^a\xi_a=0$, timelike, with eigenfields (essentially)
$\tilde{\gamma}_{ij}, \phi, Q_i, \tilde{\Gamma}^{(T)}, V^{(T)}_{ijk}$.
\item[b)] $n^a\xi_a=1$, null, with eigenfields (essentially)
$\tilde{A}^{(TT)}_{ij}$. 
\item[c)] $n^a\xi_a=(1+\frac12(7c/10-3/5-2d(2a+1)))^{1/2}\equiv C_1$, 
        with eigenfields
        (essentially)  $\tilde{A}^{(TL)}_{i}$.       
\item[d)] $n^a\xi_a= (2a/3)^{1/2}\equiv C_2$, with eigenfield (essentially) $K$
\item[e)] $n^a\xi_a= (3/5-2bc/3+16bd/3-4d/3+7c/15)^{1/2}\equiv C_3$, 
          with eigenfield (essentially) $\tilde{A}^{(LL)}$
\end{description}

In the expressions for the characteristic speeds c),d) and e),
the parameters $a,b,c,d$ are given in terms of $n_1, n_2, n_3$ via
(\ref{parametric}).  These speeds may be superluminal or causal, depending on
the values of $n_1, n_2, n_3$.  We can choose $n_1, n_2, n_3$ so that 
$C_1, C_2$
and $C_3$ are all equal to 1.  This is achieved by setting
\begin{mathletters}\label{nullens}
\begin{eqnarray}
   n_1^2 &=& \frac{4}{15}\frac{280n_2^2+49+400n_2^4}
                            {60n_2^2-49}                \\
   n_3^2 &=& \frac{6400 n_2^2}{60n_2^2-49}
\end{eqnarray}
\end{mathletters}
for any value of $n_2^2$ greater than $49/60$. This means that there is a
one-parameter family of well-posed conformally-decomposed systems with
``physical'' characteristics.  From the analytical point of view, there does not
appear to exist an argument for singling out a preferred value of $n_2$.  It is
likely that a preferred value of $n_2$ will be dictated by optimal numerical
behavior. The expressions for $a,b,c,d$ in terms of $n_2$, with $n_1$ and $n_3$
as above (\ref{nullens}), are as follows:
\begin{mathletters}
\begin{eqnarray}
    a &=& \frac32,                                      \\
    b &=& -\frac{60 n_2^2-49}{4(7+20 n_2^2)},   \\
    c &=& \frac{7}{10 n_2^2},                           \\
    d &=& -\frac{60n_2^2-49}{800n_2^2}.
\end{eqnarray}
\end{mathletters}

\subsection{Propagation of the constraints}

The propagation of the constraints can be calculated by taking a
time derivative of each one of the constraint expressions, and
subsequently using the evolution equations (\ref{combs}) to
eliminate the time derivative of the fields in the right-hand side
in favor of spatial derivatives, which recombine to yield back the
constraints.   We obtain
\begin{mathletters}\label{subs1}
\begin{eqnarray}
  \dot{\cal H}  
     &=&
         \beta^l {\cal H},_l 
        +(c-8d)\alpha e^{-4\phi}\tilde{\gamma}^{kl}{\cal P}_{k,l}
        + \cdots                \\
  \dot{\cal P}_i  
     &=&
         \beta^l {\cal P}_{i,l}
        +\frac{\alpha}{6}(1-4b){\cal H},_i
        -\frac{\alpha}{2}e^{-4\phi}
        \left(
                \tilde{\gamma}^{jl}\left(
                \tilde{\gamma}^{kr}{\cal V}_{ijk,rl}
                -\frac{7}{10}\tilde{\gamma}_{im}{\cal G}^m,_{jl}
                \right) 
                +\frac{1}{10}{\cal G}^m,_{mi}
                \right)                         \nonumber\\
                                                        &&
                -2\alpha(2a+1)e^{-4\phi}
                        \tilde{\gamma}^{jl}
                        {\cal Q}_{[i,l]j}  + \cdots             \\
        \dot{\cal G}^i  &=&
                \beta^l {\cal G}^i,_l  + \cdots                 \\
        \dot{\cal Q}_i  &=&
                \beta^l {\cal Q}_{i,l} + \cdots                 \\
        \dot{\cal V}_{ijk}      &=&
                \beta^l {\cal V}_{ijk,l}  + \cdots      
\end{eqnarray}
\end{mathletters}
where $\cdots$ denotes undifferentiated terms proportional to the
constraints themselves. 
%
%
To analyze the constraint propagation we proceed to turn (\ref{subs1}) into
first order by defining several sets of variables which represent
all the spatial derivatives of ${\cal V}_{ijk}, {\cal G}^i$ and
${\cal Q}_i$:
\begin{mathletters}
\begin{eqnarray}
   {\cal W}_{ij}        &=&
                {\cal V}_{ijk},^k
                -\frac15{\cal G}_{(i,j)}
                +\frac{1}{15}\tilde{\gamma}_{ij}{\cal G}^k,_k   \\
   {\cal X}_{ijkl} &=& {\cal V}_{ijk,l}
                -\frac13\tilde{\gamma}_{kl}{\cal V}_{ijm},^m    \\      
   {\cal U}_{ij}        &=& {\cal Q}_{[i,j]}                    \\
   {\cal Z}_{ij}        &=& {\cal Q}_{(i,j)}                    \\
   {\cal T}_{ij}        &=& {\cal G}_{[i,j]}                    \\
   {\cal J}_{ij}        &=& {\cal G}_{(i,j)}
                            +\frac{30}{7}A{\cal Q}_{(i,j)}      
\end{eqnarray}
\end{mathletters}
where $A$ is a constant which will be fixed shortly.  Calculating
the time derivative of these we obtain the resulting first-order
system of evolution of the constraints:
\begin{mathletters}\label{subs2}
\begin{eqnarray}
        \dot{\cal H}    &=&
                \beta^l {\cal H},_l 
                +\alpha(c-8d) e^{-4\phi} {\cal P}_l,^l
                + \cdots                                        \\
        \dot{\cal P}_i  &=&
                \beta^l {\cal P}_{i,l}
                +\frac{\alpha}{6}(1-4b){\cal H},_i
                -\frac{\alpha}{2}e^{-4\phi}
                        {\cal W}_{il},^l
                -2\alpha(2a+1)e^{-4\phi}
                        {\cal U}_{il},^l  \nonumber             \\
                                        &&
                -A\alpha e^{-4\phi}
                {\cal Z}_{il},^l 
                +\frac{7\alpha}{30}e^{-4\phi}
                {\cal J}_{il},^l
                +\frac{11\alpha}{30}e^{-4\phi}
                {\cal T}_{il},^l
                                                + \cdots        \\
        \dot{\cal W}_{ij}       &=&
                \beta^l {\cal W}_{ij,l}
                -\frac{c\alpha}{5} {\cal P}_{(i,j)}
                +\frac{c\alpha}{15}
                  \tilde{\gamma}_{ij}{\cal P}_l,^l
                                                + \cdots        \\
        \dot{\cal X}_{ijkl}     &=&
                \beta^m {\cal X}_{ijkl,m}
                                                + \cdots        \\
        \dot{\cal U}_{ij}       &=&
                \beta^l {\cal U}_{ij}
                +\alpha d {\cal P}_{[i,j]}      + \cdots        \\ 
        \dot{\cal Z}_{ij}       &=&
                \beta^l {\cal Z}_{ij}
                +\alpha d {\cal P}_{(i,j)}      + \cdots        \\ 
        \dot{\cal T}_{ij}       &=&
                \beta^l {\cal T}_{ij}
                +\alpha c {\cal P}_{[i,j]}      + \cdots        \\ 
        \dot{\cal J}_{ij}       &=&
                \beta^l {\cal J}_{ij}
                +\alpha \left(c+\frac{30}{7}Ad\right) 
                        {\cal P}_{(i,j)}        
                                                + \cdots        \\ 
        \dot{\cal G}^i  &=&
                \beta^l {\cal G}^i,_l  + \cdots                 \\
        \dot{\cal Q}_i  &=&
                \beta^l {\cal Q}_{i,l} + \cdots                 \\
        \dot{\cal V}_{ijk}      &=&
                \beta^l {\cal V}_{ijk,l}  + \cdots      
\end{eqnarray}
\end{mathletters}
For this system there is a symmetrizer given by
\begin{eqnarray}\label{constrenergy}
       \bar{u} \mbox{\boldmath$H$}_C u
   &=&       
        e^{4\phi}\frac{(1-4b)}{6(c-8d)} {\cal H}^2
        +                           {\cal P}_i{\cal P}^i
        +e^{-4\phi}\frac{5}{2c}     {\cal W}_{ij}{\cal W}^{ij}
        + {\cal X}_{ijkl}{\cal X}^{ijkl}
        -e^{-4\phi}\frac{2(2a+1)}{d}{\cal U}_{ij}{\cal U}^{ij}
                                        \nonumber\\
                                                &&
        -e^{-4\phi}\frac{A}{d}      {\cal Z}_{ij}{\cal Z}^{ij}
        +e^{-4\phi}\frac{11}{30c}   {\cal T}_{ij}{\cal T}^{ij}
        +e^{-4\phi}\frac{7/30}{c+Ad30/7}
                                    {\cal J}_{ij}{\cal J}^{ij} 
                                         \nonumber\\
                                                &&              
        +                           {\cal G}_i{\cal G}^i
        +                           {\cal Q}_i{\cal Q}^i
        +                           {\cal V}_{ijk}{\cal V}^{ijk}.
\end{eqnarray}
Taking $A=\frac{7}{60}(-c/d)$, $\mbox{\boldmath$H$}_C$ is positive
definite because, under the conditions
(\ref{parametric}), all the factors accompanying the squares of
the fields are strictly positive. This shows that no additional
restrictions on the ranges of the parameters $a,b,c,d$ are
necessary in order to have well posed constraint evolution.

\section{Conclusion  }\label{sec:4}


We have derived a 3-parameter family of well posed versions of the
conformally-decomposed 3+1 equations, perhaps amenable to
 successful numerical integration. 
One might object that there is no need for it in view of
the results in~\cite{BS99}, but we can argue rather strongly 
that these results may prove helpful in pinning-down the main cause of
numerical instabilities.     This well
posed version requires the lapse to be proportional to the determinant
of the intrinsic geometry of the surfaces, and requires combinations
of the contraints with the evolution equations. The lapse density
$\sigma$ and the shift vector $\beta^i$ are arbitrary non-dynamical
variables, which means that they  must be specified as free source
functions. This well posed version uses the same variables as the
original system (except for the addition of the first spatial
derivatives of the densitized 3-metric, referred to as ``conformal
metric'' by the authors in~\cite{BS99}).  In addition, this well-posed
version of the original equations propagates the constraints in a
stable manner, which is relevant to unconstrained evolution. We think
that this is the least invasive way to turn the original
conformally-decomposed system into a well posed one.  In practice, a
choice of the numerical parameters $n_1, n_2, n_3$ must be made. The
characteristic speeds depend on this choice.  

Optimal choices of the parameters $n_1, n_2, n_3$ for numerical
evolution are those that ensure that the characteristics are all
either null or timelike.  With such a choice, the formulation would be
suited to evolve blackhole spacetimes outside the event horizon. Among
these choices, it has been suggested~\cite{fixing} that the preferred
one would be the one for which the characteristics are all
``physical'', namely, either null or normal to the slices.  We have shown
that such a choice is possible for an arbitrary $n_2>\sqrt{49/60}$.

The systems obtained in this work are not contained in our
previous work~\cite{frittelliletter,frittelli-reula99}. The choice
of variables in~\cite{frittelliletter,frittelli-reula99} is
inadequate for decomposing the trace and trace free part of the
extrinsic curvature, as well as for extracting the determinant of
the 3-metric.  This is clear from the fact that, in that work, the
available parameters $\alpha$ and $\beta$ are not allowed to take
the value $-1/3$ without the argument breaking down.  

The systems obtained here differ significantly from the system
obtained in~\cite{brugman} by considering the trace of the
extrisic curvature $K$ and the determinant of the instrinsic
metric (and its derivatives) as dynamical variables on equal
footing with the rest, rather than as free source functions.
Furthermore, we have obtained a 3-parameter family of systems, 
one system for each
appropriate choice of $n_1, n_2, n_3$, whereas
in~\cite{brugman} there is only one system which preserves the
trace conditions. Additionally, we have separated the divergence
of the intrinsic metric $\tilde{\Gamma}^i$ from the
divergence-free part of the metric. This decomposition keeps up
with the spirit of~\cite{BS99}.

We have found that in obtaining these well-posed formulations the
lapse must be proportional to some power of the determinant of the
intrinsic metric, since  the parameter $a$ cannot take the
value
0.
This is similar to our findings
in~\cite{frittelli-reula94,frittelliletter,frittelli-reula99}, as
well as other notable cases~\cite{helmut96,mirta97,mirta98,york98}.  
In our present case this is remarkable, since we have used quite
general energy norms.

\acknowledgments

We thank Roberto G\'{o}mez for enlightening 
discussions. This work
has been supported by NSF under grant No. PHY-9803301, and by CONICOR, CONICET,
and SeCyT, UNC. 


\begin{thebibliography}{10}

\bibitem{bona-masso92}
C. Bona and J. Masso, Phys. Rev. Lett. {\bf 68},  1097  (1992).

\bibitem{frittelli-reula94}
S. Frittelli and O.~A. Reula, Commun. Math. Phys. {\bf 166},  221 
(1994).

\bibitem{massoletter}
C. Bona, J. Masso, E. Seidel, and J. Stela, Phys. Rev. Lett. {\bf
75},  600
  (1995).

\bibitem{york}
A. Abrahams, A. Anderson, Y. Choquet-Bruhat, and J.~W. York, Phys.
Rev. Lett.
  {\bf 75},  3377  (1995).

\bibitem{frittelliletter}
S. Frittelli and O.~A. Reula, Phys. Rev. Lett. {\bf 76},  4667 
(1996).

\bibitem{helmut96}
H. Friedrich, Class. Quantum Grav. {\bf 13},  1451  (1996).

\bibitem{frittelli-reula99}
S. Frittelli and O.~A. Reula, Well posed first-order forms ot eh
3+1 Einstein
  equations, in preparation (January 1999).

\bibitem{kreissbook}
B. Gustaffson, H.-O. Kreiss, and J. Oliger, {\em Time-dependent
problems and
  difference methods} (Wiley, New York, 1995).

\bibitem{kurki-laguna-matzner93}
H. Kurki-Suonio, P. Laguna, and R.~A. Matzner, Phys. Rev. D {\bf
48},  3611
  (1993).

\bibitem{shibata}
M. Shibata and T. Nakamura, Phys. Rev. D {\bf 52},  5428  (1995).

\bibitem{BS99}
T.~W. Baumgarte and S.~L. Shapiro, On the numerical integration of
Einstein's
  field equations, gr-qc/9810065, to appear in Phys. Rev. D.

\bibitem{lambdavar}
O. Brodbeck, S. Frittelli, P. H\"{u}bner, and O.~A. Reula, J. Math.
Phys. {\bf
  40},  909  (1999).

\bibitem{helmutnagy}
H. Friedrich and G.~B. Nagy, The initial boundary value problem for
Einstein's
  field equations, 1998, ~AEI-preprint-078.

\bibitem{stewart98}
J.~M. Stewart, Class. Quantum Grav. {\bf 15},  2865  (1998).

\bibitem{brugman}
M. Alcubierre, B. Brugmann, M. Miller, and W.-M. Suen, A conformal
hyperbolic
  formulation of the {E}instein equations, preprint, gr-qc/9903030.

\bibitem{york79}
J.~W. York,  in {\em Sources of Gravitational Radiation} (Cambridge
University
  Press, Cambridge, 1979).

\bibitem{choquet-ruggeri83}
Y. Choquet-Bruhat and T. Ruggeri, Commun. Math. Phys. {\bf 89}, 
269  (1983).

\bibitem{york98}
A. Anderson and J. James W.~York, Phys. Rev. Lett. {\bf 81},  1154 
(1998).

\bibitem{fixing}
A. Anderson and J. James W.~York, Fixing Einstein's equations,
preprint,
  gr-qc/9901021.

\bibitem{mirta97}
  M. Iriondo, E. Leguizam\'on and O. Reula,
  Phys. Rev. Lett. {\bf 79} 4732-4735  (1997)

\bibitem{mirta98}
  M. Iriondo, E. Leguizam\'on and O. Reula,
  Adv. Theor. Math. Phys. {\bf 2} 1075-1103 (1998)
\end{thebibliography}

\end{document}